\documentclass[conference]{IEEEtran}
\IEEEoverridecommandlockouts
\usepackage{cite}
\usepackage{amsmath,amssymb,amsfonts}
\usepackage{algorithmic}
\usepackage{graphicx}
\usepackage{textcomp}
\usepackage{xcolor}
\def\BibTeX{{\rm B\kern-.05em{\sc i\kern-.025em b}\kern-.08em
    T\kern-.1667em\lower.7ex\hbox{E}\kern-.125emX}}
\begin{document}

\title{Implementation of OpenAirInterface-based real-world channel measurement for evaluating wireless transmission algorithms\\
%{\footnotesize \textsuperscript{*}Note: Sub-titles are not captured in Xplore and
%should not be used}
%\thanks{Identify applicable funding agency here. If none, delete this.}
}

\IEEEpeerreviewmaketitle
\author{\IEEEauthorblockN{Qiuheng Zhou, Wei Jiang}
	\IEEEauthorblockA{\textit{German Research Center for Artiﬁcial Intelligence (DFKI GmbH), Kaiserslautern, Germany} \\
		Emails: $\{$qiuheng.zhou, wei.jiang$\}$@dfki.de\\
}
		}

\maketitle

\begin{abstract}
The fourth-generation Wireless Technology (4G) has been adopted by all major operators in the world and has already ruled the cellular landscape for around a decade. A lot of researches and new technologies are being considered as potential elements contributing to the next generation wireless communication (5G). The lack of realistic and flexible experimentation platforms for collecting real communication data has limited and slowed the landing of new approaches. Software Defined Radio (SDR) can provide flexible, upgradable, and long lifetime radio equipment for the wireless communications infrastructure, which can also provide more flexible and possibly cheaper multi-standard-terminals for end users. By altering the open-source code individually, we can freely perform the real value measurement. This paper provides a real Long Term Evolution (LTE) channel measurement method based on the OpenAirInterface (OAI) for the evaluation of the channel prediction algorithm. Firstly, the experimentation platform will be established by using OAI, Universal Software Radio Peripheral (USRP), and commercial User Equipment (UE). Then, some source codes of OAI are analyzed and changed, so that the real-time over-the-air channel measurement can be achieved. The results from the measurement are then trained and tested on the channel prediction algorithm. The results of the test illustrate that the implemented channel measurement method can meet the need for algorithms' verification and can be further extended for more development of algorithms.
\end{abstract}


\section{Introduction}
As we all know, 5G is finally becoming a reality. Researches for the next-generation mobile networks have already begun with the examination and evaluation of candidate technologies and architectures. Among them, researches are done, which utilize artificial intelligent methods, that process the outdated Channel State Information (CSI) to forecast the future CSI, so that the channel predictor can achieve a very high prediction accuracy in a simulated fast fading channel without any prior knowledge \cite{b1,b2,b3}. However, performance assessment over wireless transmission technologies requires strict evaluation and real-world validation before deployment. While software for over-the-air simulation has evolved significantly over the years, it still cannot capture the complex real-world environment completely. Real-world evaluation over platforms with commercial LTE equipment, however, is restricted in individual configuration capabilities, mainly because of commercial considerations. This has resulted in the need for an open-source experimentation platform with a high degree of flexibility, that the researchers can commonly use for understanding the complexities associated with real-world settings while at the same time obtain reproducible and verifiable results. SDR system has a great advantage in simulating and verifying the new communication technologies attributed to its striking advantage in ﬂexibility and reconﬁgurability. A typical General Purpose Processor (GPP) based SDR system consists of a single piece of peripheral equipment connecting to the GPP. The peripheral equipment, e.g., USRP, is mainly responsible for frequency conversion and digitization. The GPP is responsible for baseband signal processing \cite{b4}. OAI \cite{b5} is fully open-source and provides a complete software implementation of the entire LTE system architecture on GPP, which is more appealing and appropriate for developing and evaluating our proposed algorithm. OAI can verify various new communication technologies by changing some code but not the hardware. To provide real wireless communication CSI for verifying the proposed channel prediction algorithms \cite{b1,b2,b3} and even for some other use cases later, a real LTE daily connection scenario between the LTE base station and a smartphone is implemented so that we can perform uplink wireless channel measurement by analyzing and changing the source codes from OAI, e.g., uplink channel estimation and uplink scheduling.
\par The rest of this paper is organized as follows: Section \ref{section2} introduces the system platform for channel measurement. Section \ref{section3} provides the idea about how we perform channel measurement. Section \ref{section4} presents the result of the measurement and the verification of the proposed channel prediction algorithm. Finally, Section \ref{section5} remarks on this paper.

\section{System introduction}\label{section2}
	\begin{figure}[htbp]
	\centering
	\includegraphics[width=\linewidth]{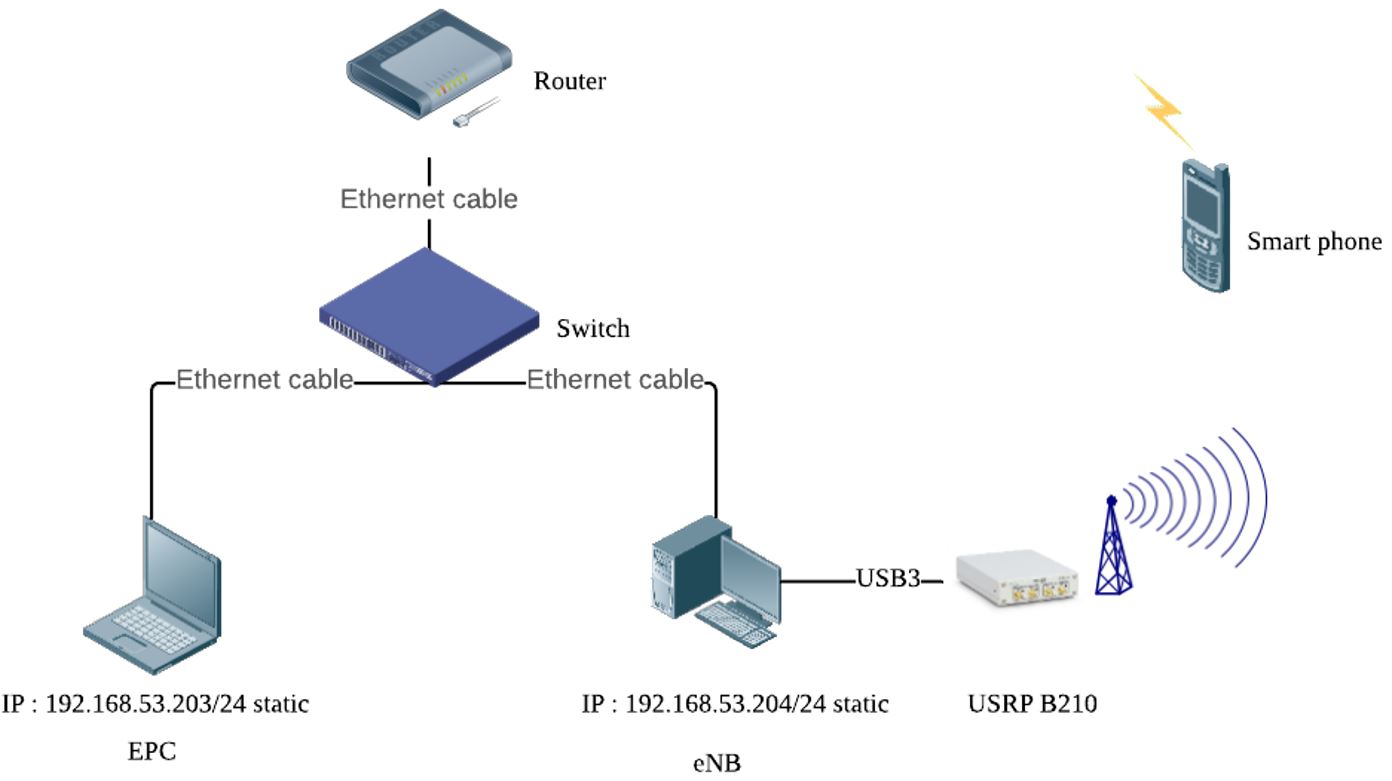}
	\caption{The hardware setup of LTE communication platform}
	\label{fig1}
	\end{figure} 
The system setup can be seen in Fig.~\ref{fig1}. There are two computers for running OAI Evolved Node B (eNB) and OAI Evolved Packet Core (EPC) respectively. Here the Intel (R) Xeon (R) CPU E3-1245 with 3.7 GHz and Intel (R) i5-5300U with 2.3 GHz are selected for each. According to our working environment, the two computers are connected to the switch. One thing that should be paid attention, that the Dynamic Host Conﬁguration Protocol (DHCP) from the switch to allocate IP addresses is not used, instead the static IP for the computers is set. The computer, which runs OAI eNB, should have a USB 3.0 port for communication with USRP. The computers with static IP addresses can ping to each other through the switch, at the same time have access to the Internet through the switch, which connects to a router. The USRP receives the signal from the eNB through a USB 3.0 interface and broadcasts the signal to the air. The conﬁgured smartphone can then access the signal and connect to this network.
\par
\begin{figure}[htbp]
	\centering
	\includegraphics[width=\linewidth]{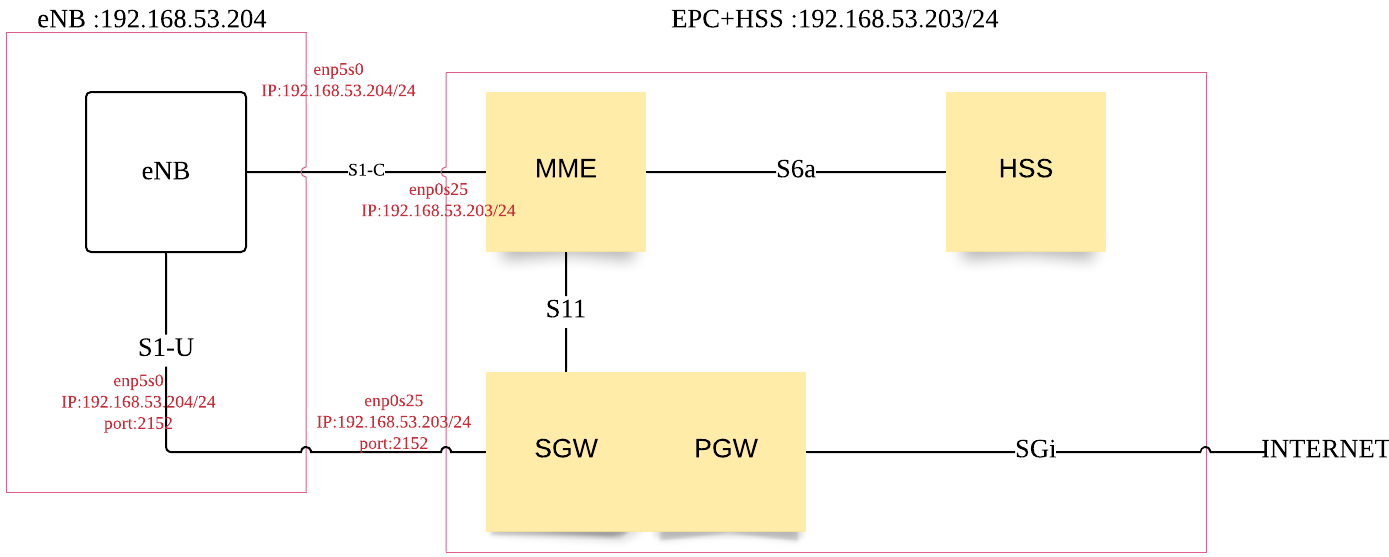}
	\caption{The IP setup of the LTE platform}
	\label{fig2}
	\end{figure}

 Referred to the OAI architecture in \cite{b5}, the hardware setup and the OAI interior architecture and interfaces are combined to assign the IP address to each interface, which will then be used in the conﬁguration from installation. In Fig.~\ref{fig2}, the IP assignments for the interfaces are depicted. There are two planes of the S1 interface between the OAI eNB and OAI EPC, respectively are the user plane (S1-U) and the control plane (S1-C). The S1 interface in LTE is used between eNBs and the EPC, speciﬁcally, the Mobility Management Entity (MME) and Serving Gateway (SGW). The MME and SGW will share the IP address of EPC to connect to the eNB and then connect with each other through the S11 interface, which is based on GTP-control with some additional functions for paging coordinating. The other interfaces are already set in OAI as default. 
\par Followed by the EURECOM OAI community tutorials \cite{b6}, the LTE communication platform can be established step by step, including the configuration and installation of eNB, USRP and EPC, UE configuration, USIM card programming, and user ID registration on the HSS database. Huge attention is paid to avoid false settings from detail so that we can finally get a stable LTE communication.
%as shown in Fig.~\ref{fig3}, which real-timely monitors the communication states and logs.
%	\begin{figure}[htbp]
%	\centering
%	\includegraphics[width=\linewidth]{enb_with_ue.png}
%	\caption{OAI softmodem monitor with UE connected}
%	\label{fig3}
%	\end{figure} 

\section{Channel measurement}\label{section3}
Based on the hardware setup in Fig.~\ref{fig1}, it is more realistic that the channel measurement is performed on the side of eNB, which means, the eNB acts as a receiver to get the uplink data from the UE and estimate the uplink channel. In OAI, all uplink channels are available, including Physical Random Access Channel (PRACH), Physical Uplink Shared Channel (PUSCH), Physical Uplink Control Channel (PUCCH), Sounding Reference Signal (SRS), and Demodulation Reference Signal (DRS). So, all channels can be configured by changing the fully opened source code. For verifying the channel prediction algorithm in this paper, the channel estimation is mainly focused. Channel prediction helps to improve the performance of an operation on both sides of the transceiver and receiver. The channel prediction acts as a middleware for processing the channel estimates so that the channel can be better decoded and react to the channel emergency previously. That means the more frequently the channel estimates are measured, the better the channel prediction will perform. 
\par
\begin{figure*}[htbp]
		\centering
		\includegraphics[width=\linewidth]{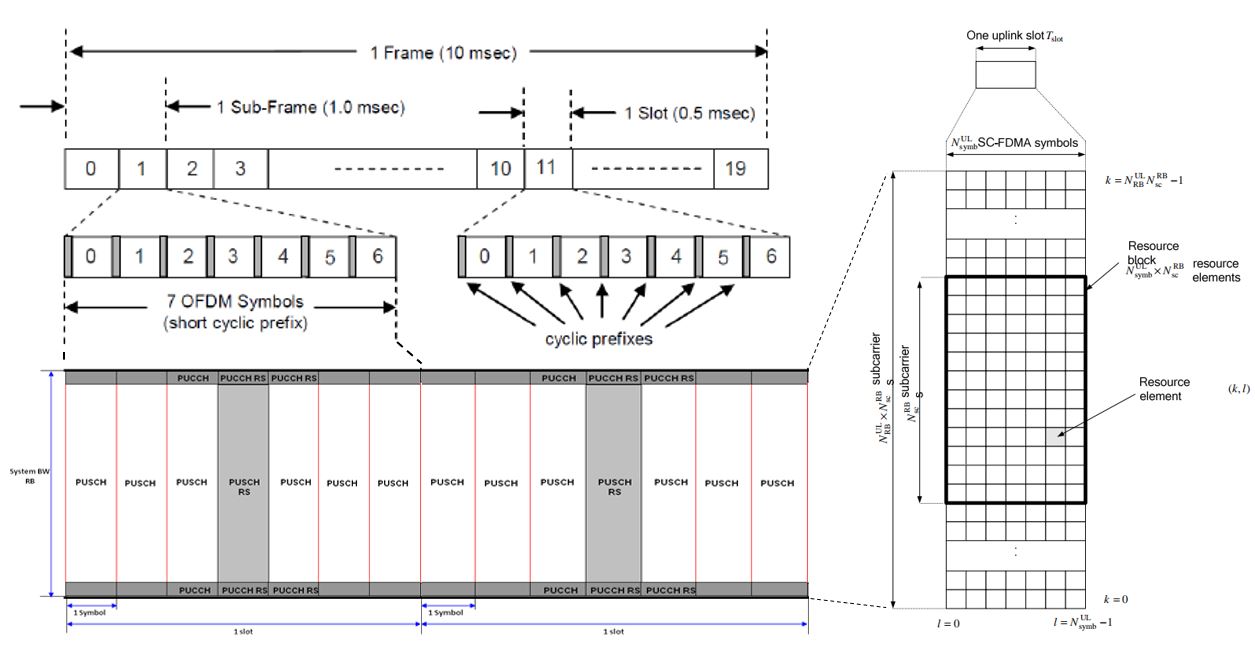}
		\caption{LTE FDD frame structure format with uplink resource grid map}
		\label{fig3}
	\end{figure*}

 On the basis of 3GPPP LTE standards \cite{b7}, in the FDD system, each radio frame is 10 ms long and consists of 10 subframes. There are 14 OFDM symbols in one subframe. As shown in Fig.~\ref{fig3}, 2 slots are contained in one subframe. Normally, the reference signals are carried on the fourth OFDM symbol position in each time slot, where the channel estimation \cite{b8} takes place.

\par
\begin{figure}[htbp]
	\centering
	\includegraphics[width=\linewidth]{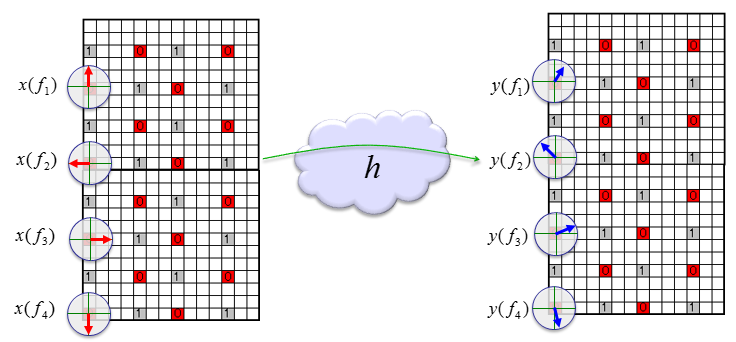}
	\caption{Channel estimation by comparing transmitted and received reference signal}
	\label{fig4}
	\end{figure}

 There are many different ways for channel estimation, but fundamental concepts are similar. The process is done as follows:
\begin{itemize}
    \item Set a mathematical model to correlate transmitted signals and received signals using the channel matrix.
    \item Transmit a known signal (we normally called this a reference signal or pilot signal) and detect the received signal.
    \item By comparing the transmitted signal and the received signal, each element of the channel matrix can be figured out.
\end{itemize}
As presented in Fig.~\ref{fig4}, under a Single-Input Single-Output (SISO) situation, reference signals are embedded onto only one antenna port. The vertical line in the downlink resource grid map in Fig.~\ref{fig4} represents the frequency domain. The reference signals are indexed with $f_1$, $f_2$, $f_3$...$f_n$. Each reference symbol can be a complex number (In-phase and Quadrature (I/Q) data) that can be plotted as shown in Fig.~\ref{fig4} (red and blue arrows). Each complex number (Reference Symbol) on the left (transmission side) is modiﬁed (distorted) to each corresponding symbol on the right (received symbol). Channel Estimation is the process of ﬁnding a correlation between the array of complex numbers on the left and the array of complex numbers on the right. The detailed method of the channel estimation can very be depending on the implementation. Since our platform will ﬁrstly use SISO, then the system model for each transmitted reference signal and received reference signal can be represented as follows. $y()$ represents the array of the received reference signal, $x()$ represents the array of the transmitted reference signal, and $h()$ represents the array of channel coefﬁcient. $f_n$ is the indices.
\begin{equation}\label{eq1}
    y(f_n)=h(f_n) \cdot x(f_n)
\end{equation}
The $x()$ is known because it is given from the transmitted signals. And the $y()$ is also known because it is measured/detected from the receiver. With this information, the channel coefficient array can be calculated as shown in equation \ref{eq2}, where $H$ is a symbol, which represents the Hermitian of a matrix.
\begin{equation}\label{eq2}
    h(f_n)= y(f_n) \cdot x^H(f_n)
\end{equation}
After this channel coefficient estimation process on the reference signal symbol position, most of the applications will also continue to estimate all the channel coefficient at all locations, where there is no reference signal. In this paper, only the original channel estimates for testing the channel prediction algorithm are needed, that we directly read all the estimated channel coefficients after the channel estimation is done on every reference symbol position.
\par
\begin{figure}[htbp]
	\centering
	\includegraphics[width=\linewidth]{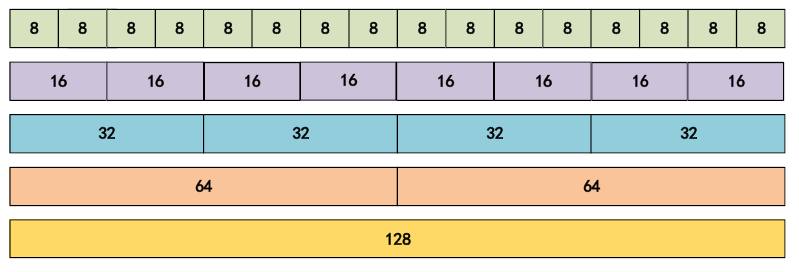}
	\caption{Data storage format of 128 bit data stream}
	\label{fig5}
	\end{figure}

 On the right side of Fig.~\ref{fig3}, the resource grid map is depicted. The symbols are indexed with $l$ and the subcarriers are indexed with $k$. The Resource Elements (REs) are located at $(k,l)$ coordinates. There are 12 REs in a Resource Block (RB). So, if the 5MHz bandwidth is used here for transmission, then there will be 25 RBs in the frequency domain of the uplink channel. In OAI, the channel estimates are calculated by using the Streaming Single Instruction Multiple Data Extensions (SSE) \cite{b9} structure on each reference signal symbol position. A common usage of the SSE is the complex value multiplication. Each RE consists of 2*16 bit value (real and imaginary). As seen in Fig.~\ref{fig5}, in each thread of data stream, the 128-bit data stream includes 4 REs. The channel estimates are calculated by multiplying the complex conjugate of the received reference signal and the receiver generated reference signal every time slot. So, under the situation of 5MHz transmission, there can be up to 25RBs allocated for channel estimation, which will need $25\times12\times2\times16 bit / 0.5 ms=19.2Mbit/s$ data transmission speed. For getting the channel estimates continuously on every resource element of every reference symbol position, the read/write speed should be taken into consideration.
OAI has provided a GUI for real-timely monitoring of the communication states and logs. With the help of the monitoring GUI, all sorts of information can be displayed. Then the channel estimates can be directly output on the Linux terminal command window, which has the data transmission between the processor and memory with nono-speed level I/O speed so that it can display various system status information without any impact on the real-time computation.
\par Although the channel estimates can real-timely be output, there is still a challenge for configuring the system, so that it can keep a constant bandwidth for uplink transmission for a certain UE and also make sure that the channel estimation happens continuously on every reference signal position. In a random access network, to dynamically and ﬂexibly arrange the best channel for active UEs, the users' Quality of Service (QoS) is ranked by the receiver eNB. Commonly, the signal-noise ratio is converted to be the channel quality index, and the ranked channel qualities are referenced from the scheduler so that the subcarriers in the frequency domain and the subframes in the time domain can be allocated for a certain uplink channel demodulation, which aims at limiting the network's congestion and enhancing the service quality \cite{b10}. For verifying a channel prediction algorithm, however, the scheduling and dynamic RB allocation should be disabled on the contrary. Since there is only one UE in a SISO system, actions are that the uplink channel quality index, which is represented by SNR, is then set constant, if the value of SNR located itself in a reasonable range for normal transmission. So that the UE receives a constant power index for the RB allocation, which ensures a contiguous uplink transmission with constant bandwidth.

\section{Experimental results and analysis}\label{section4}
\begin{figure}[htbp]
	\centering
	\includegraphics[width=\linewidth]{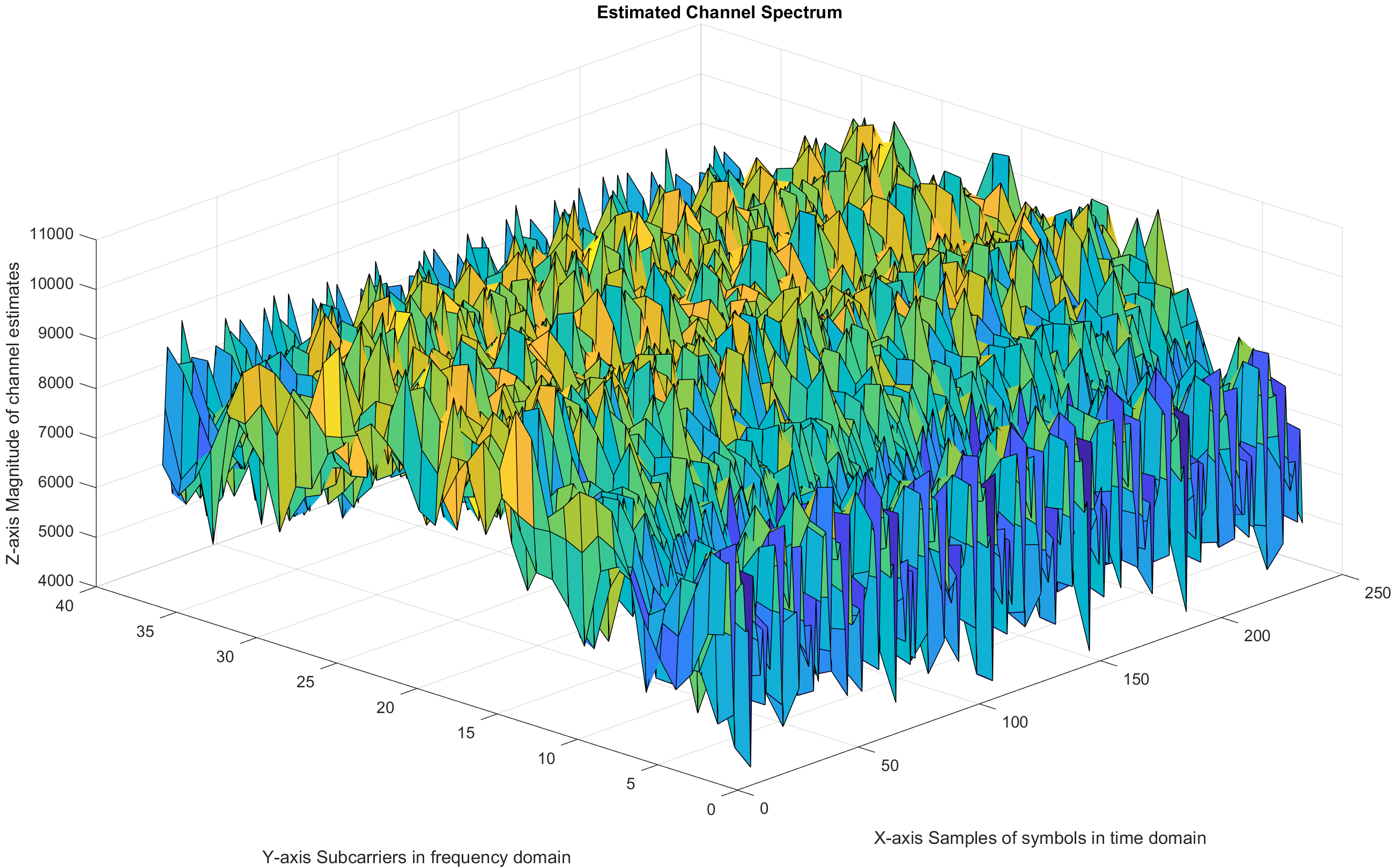}
	\caption{Estimated Channel Spectrum}
	\label{fig6}
	\end{figure}

After having changed the source codes from OAI, rebuild the eNB, and reboot the system, the channel estimates are extracted and formatted, so that it can be readable and visualized in MATLAB. Since there are only integers from real and imaginary parts that are extracted, the massive integers should be converted back to be the complex values in the frame structure. Fig.~\ref{fig6} has shown some samples of the spectrum from our measurement. In this measurement, the allocated bandwidth of the uplink channel is constrained to 3 RBs, which are 36 subcarriers on the Y-axis. So, there is much less pressure for the processor to calculate channel estimates during a time symbol. There are 240 time symbols shown here on the X-axis, which represents 0.12s duration in the time domain. The magnitude of the complex-valued channel estimates of each resource element is then plotted on the Z-axis.
\par
\begin{figure}[htbp]
	\centering
	\includegraphics[width=\linewidth]{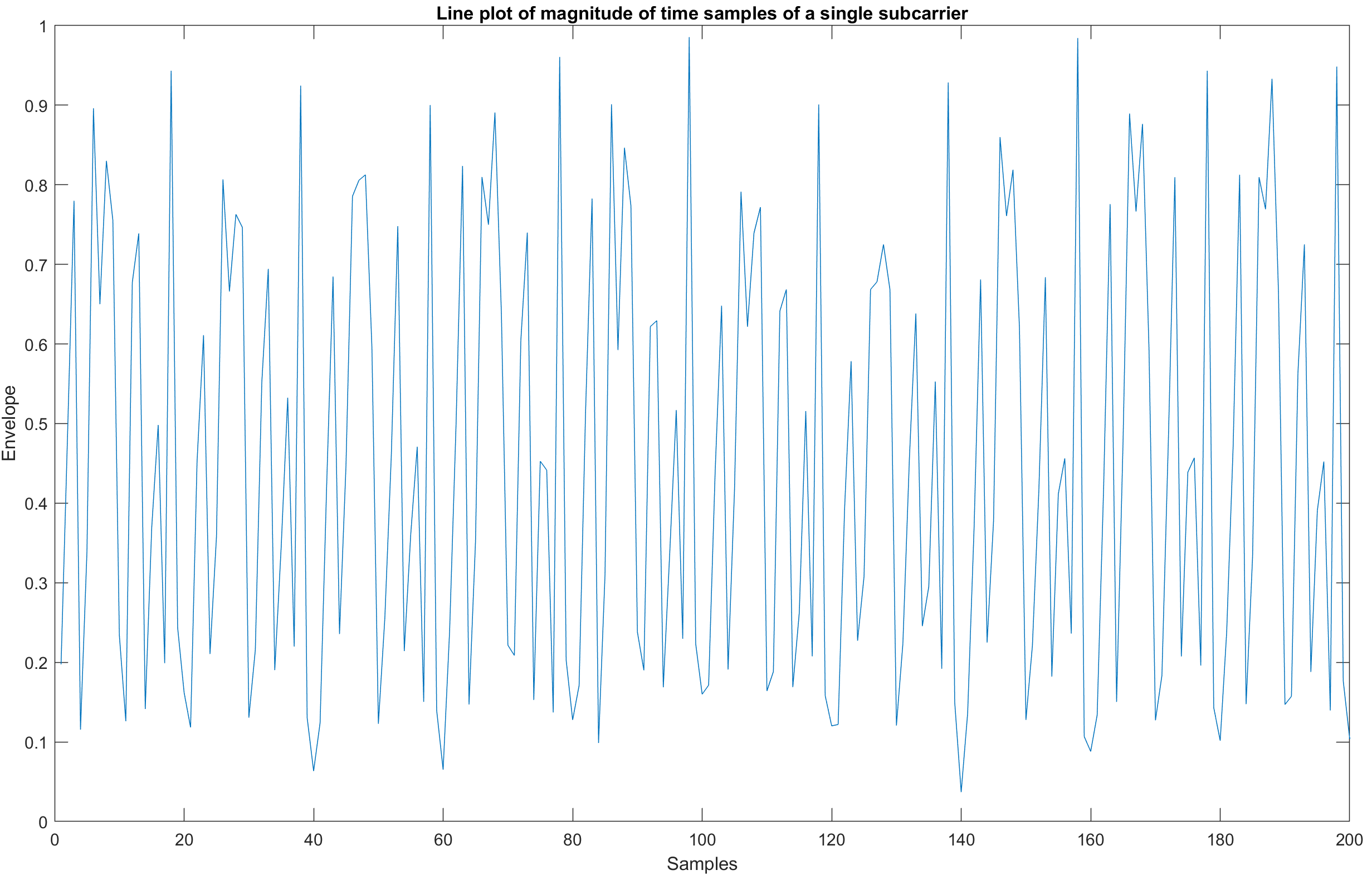}
	\caption{Estimated channel spectrum of a single subcarrier}
	\label{fig7}
	\end{figure}

 A random subcarrier of the frequency domain is chosen and the magnitude of the 200 channel estimates from the 240 time symbols are then normalized and plotted, which can be seen in Fig.~\ref{fig7}. Remember that the channel estimates are extracted on every reference signal symbol position in each slot of a frame, which means every 7 symbols in the time domain, so, the plots cannot show a smooth curve like the simulated channel results in \cite{b1}.
\par
\begin{figure}[htbp]
	\centering
	\includegraphics[width=\linewidth]{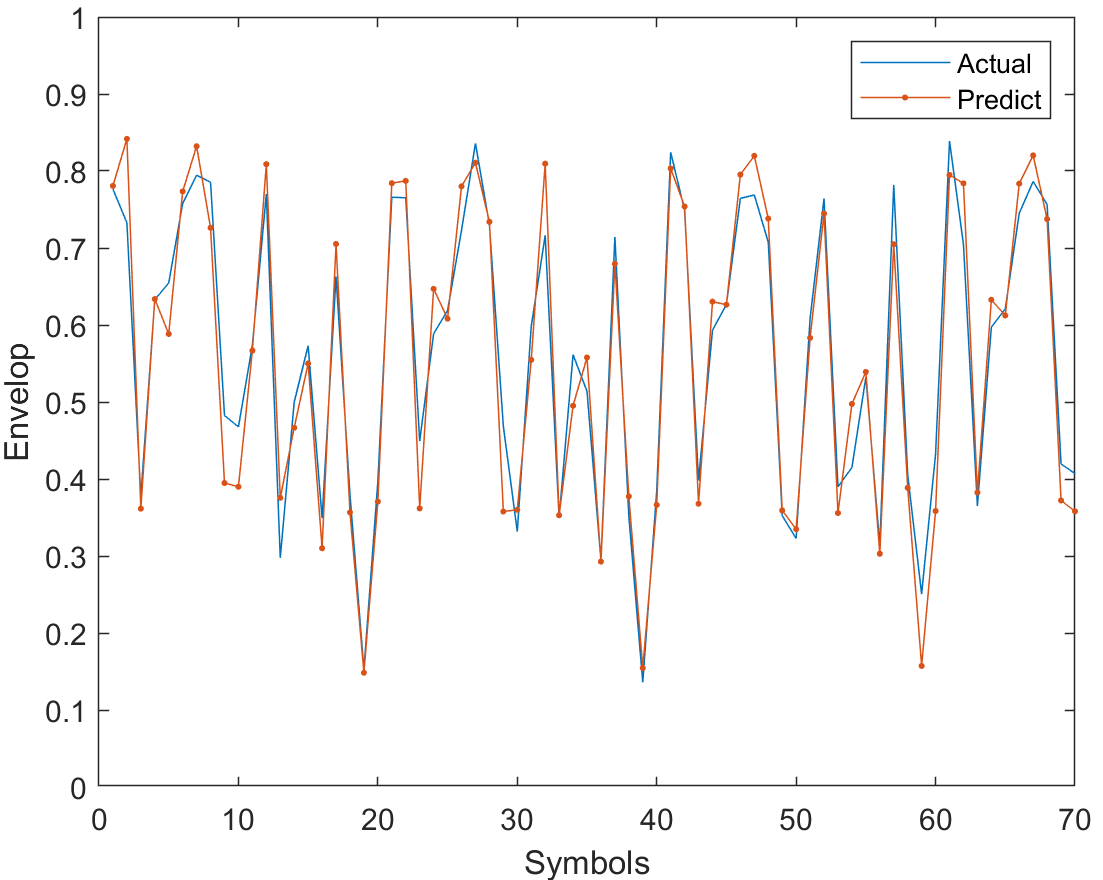}
	\caption{Channel prediction on the envelope of real-world estimated channel}
	\label{fig8}
	\end{figure}

 In \cite{b1}, the author has normalized the envelope and the real part of the complex value of the channel estimates using Z-score normalization because the channel is assumed to be a flat-fading Rayleigh channel with an average power gain of $0dB$, where the channel coefficient $h$ is zero-mean circularly-symmetric complex Gaussian random variable with the variance of 1, i.e., $h \sim \mathcal{C}\mathcal{N}(0,1)$. The results of this paper have shown a great performance of the channel prediction algorithm. However, it lacks the prediction results on the real-world communication coefficients. In this paper, the channel estimates are predicted with an interval of 7 time symbols. The algorithm is then learned and tested on the measured channel estimates. The result is shown in Fig.~\ref{fig8}, which has an accuracy of Mean Square Error (MSE) $\sigma^2=2.3\times10^{-3}$. Compared to the channel prediction results in \cite{b1}, which has a higher accuracy with $\sigma^2=4.97\times10^{-6}$ on the case of one-symbol-ahead and $\sigma^2=1.29\times10^{-4}$ on the case of 5-symbols-ahead, the accuracy of the prediction has verified the advancement of this algorithm, which also proves that the proposed channel measurement method is feasible.
	
\section{Conclusion}\label{section5}
This paper has aimed to build a real-world open-source SDR application-based LTE-access experimentation platform for performing the channel measurement so that we can verify the AI-based next-generation wireless transmission algorithms. All the set tasks have been perfectly met. A stable LTE communication platform is built, which supports the later configured channel estimates measurement use case. In consideration of our platform setup, we have chosen the uplink channel for research. By changing the uplink channel schedule referred to the 3GPPP standards, the bandwidth of the uplink channel can be restrained and made constant, which leads to a successful channel estimates measurement every 7 symbols in the time domain. The normalized measurement data was then tested on the proposed channel prediction algorithm, which has shown an acceptable accuracy on channel prediction. It is the first approach that the OAI has been used for directly retrieve the channel estimates consecutive in every time slot, which provides more frequent measurement result. Our implemented OAI and USRP-based experimentation platform has been proved to be powerful enough for some research and veriﬁcation purposes, which can also be extended for other tasks.

\end{document}